\documentstyle[twoside,fleqn,jhuwkshp,psfig]{article}
\newcommand{\etal}{{\it et al.} }

\begin{document}

\title{TRANSIENT PHENOMENA AND OUTBURSTS FROM GALACTIC BLACK-HOLE CANDIDATES}
 
\author{T. Belloni
\address{Osservatorio Astronomico di Brera, 
Via E. Bianchi 46, I-23807 Merate (LC), Italy}}

\begin{abstract}
The RossiXTE mission provided us with an unprecedentedly large database
of X-ray observations of transient black-hole candidates. These systems
are crucial for the understanding of the physical properties of mass
accretion onto black holes. Here I
review the results on a selected sample of systems and describe their
behavior in a purely phenomenological way. From these results, we
can derive a better classification of the spectral and timing characteristics
of black hole candidates in terms of basic states.

\end{abstract}

\maketitle

\section{Before RXTE: spectral/timing states}

Transient black hole candidates (TBHC) are the most important laboratories
where we can study accretion onto black holes. The main reasons are two:
first, there is a very limited number of bright persistent systems; second,
transients go through a large range of mass accretion rate during their
outbursts, therefore allowing us to study how the accretion properties
change with accretion rate. In the case of persistent sources, they show
rare state transitions (if any), and even when they do it is not clear what
their dependence on accretion rate is (see e.g. \cite{zhan1997}).

Before the launch of the Rossi X-ray Timing Explorer (RXTE), a relatively
small number of systems was known (see \cite{tana1995,vdk1995}
for a review). From these sources, a general ``canonical'' paradigm
for the spectral/timing properties of TBHC had emerged 
(see \cite{tana1995,vdk1995,miya1993}). Although there were
notable exceptions, this paradigm was a good starting point for theoretical
modeling. Four separate states were identified, which will be
described here according to the behavior in the 2-20 keV band:
\begin{itemize}

\item Low/Hard State (LS): the energy spectrum can be described by a single
	power law with a photon index $\Gamma\sim$1.6. In addition, sometimes
	a weak disk-blackbody (DBB) component with kT$<$1 keV is
	observed, contributing less than a few percent to the detected
	flux. The Power Density Spectrum (PDS) is characterized by a strong
	band-limited noise with a break frequency below 1 Hz and a 
	large fractional variability (30-50\%). A low-frequency $<$1 Hz QPO is
	sometimes observed.

\item Intermediate State (IS): the energy spectrum can be decomposed in 
	two components: a power law with $\Gamma\sim$2.5 and a clearly
	detectable DBB with kT$\leq$1 keV. The PDS shows a band-limited
	noise with a break frequency higher than the LS (1-10 Hz) and
	a fractional variability of 5-20\%. Sometimes a 1-10 Hz QPO is
	observed.

\item High/Soft State (HS): the energy spectrum is dominated by a DBB
	component with kT$\sim$1 KeV, with the power-law component either
	below detection or extremely weak and steep ($\Gamma\sim$ 2--3). Very
	weak noise is detected in the PDS, in the form of a power-law
	with a few \% of fractional variability.

\item Very High State (VHS): the energy spectrum is a combination of a
	DBB (kT$\sim$1-2 keV) and a power law ($\Gamma\sim$ 2.5).
	The PDS can be of two types: either a band-limited noise
	similar to that of the IS, or a power-law, sometimes with a
	QPO. This state was observed only in two sources: the transient
	GS 1124-68 and the persistent source GX 339-4.

\end{itemize}

The dependence of these states and their transitions on increasing 
accretion rate
was determined mostly by the only transient system that had shown all
four of them, GS 1124-68 \cite{miya1994,ebis1994}: LS--IS--HS--VHS.
As one can see from the description above, the IS is very similar to the 
VHS, both in energy and timing characteristics. What was believed to 
be different between them is the value of the accretion rate: the VHS
was observed at very high accretion rates, while the IS was observed
much later in the outburst of GS 1124-68, after a long period of HS, and
therefore at lower accretion rate (see \cite{bell1997}).

Despite the exceptions, a transient was expected to have a 
fast-rise/exponential-decay light curve (reflecting the time history
of accretion rate), lasting a few months, possibly
with one or more re-flares, undergo a number of state transitions in the
sequence outlined above following accretion rate changes, and return to 
quiescence after a period of a few weeks to months, until the next outburst.
Once again, the prototypical source would be GS 1124-68.

\section{The RXTE ``outburst''}

With the advent of RXTE at the end of 1995 \cite{brad1993,swan1998},
our view of TBHC has been
changed and widened. Thanks to the All-Sky Monitor (ASM) on board RXTE
\cite{levi1996},
we have been able to discover promptly both new transients and new outbursts
of known ones. The flexibility of the mission allowed the accumulation 
of a vaste and valuable database of pointed observations with the narrow
field instruments on board, covering the wide range 2-200 keV. In
particular, the large collecting area of the PCA instrument 
\cite{zhan1993,jaho1996} yielded
high signal-to-noise data for timing and spectral analysis, and the
availability of the RXTE archive on-line made these datasets easily
accessible to the community. This resulted in a real ``outburst'' both
of new sources and of publications on the subject. 

In the rest of the paper, I will concentrate on the results on a few
major systems and show how these have changed our overall picture. My
approach will be mostly on the phenomenological side.

\section{XTE J1550-564}

\begin{figure}[t] 
\vspace{10pt}
\centerline{\psfig{file=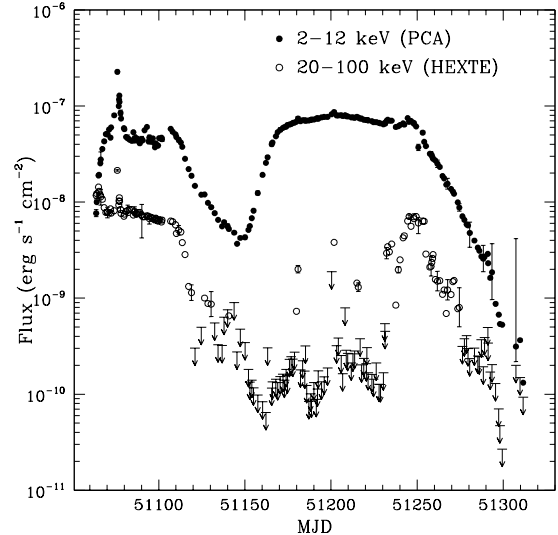,width=3.0in}}
\caption{PCA and HEXTE light curves of the 1998/1999 outburst of XTE J1550-564
from \cite{sobc2000}.
}\label{fig:sobczak6}
\end{figure}

This transient was discovered in September 1998 with RXTE and its outburst
lasted until May 1999 (see \cite{sobc2000} and references therein). Two more
outbursts were detected in 2000 and 2001, but I will concentrate on the
1998/1999 one. As one can see from Fig. \ref{fig:sobczak6}, 
the light curves in the PCA
and HEXTE instruments look quite different, indicating large spectral 
changes.

\begin{figure*}[t] 
\vspace{10pt}
{\psfig{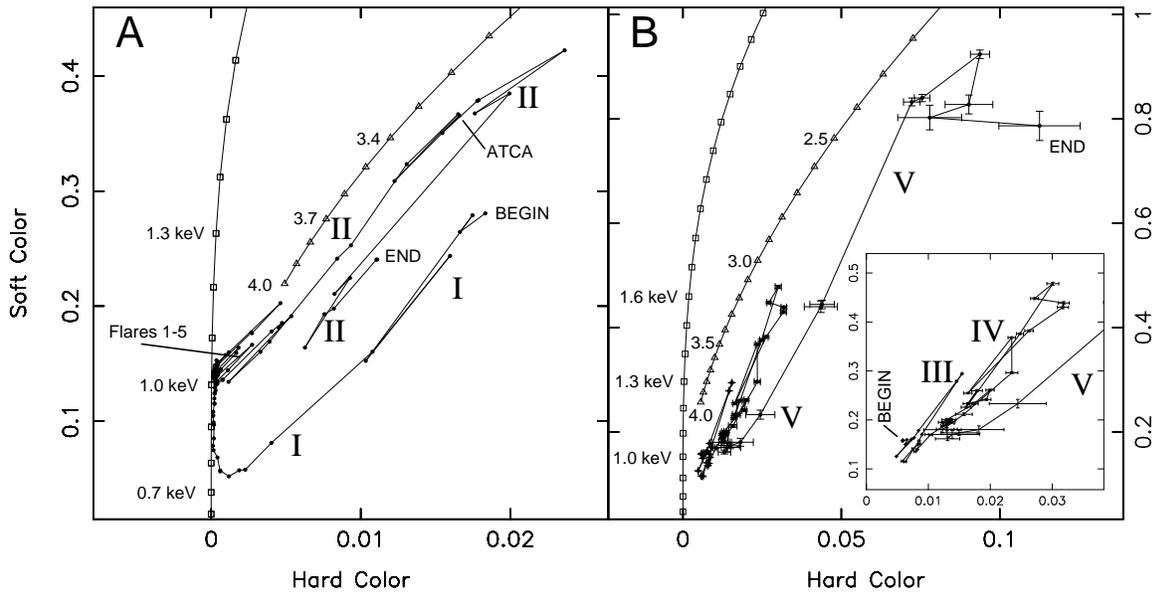}}
\caption{Color-color diagrams for the second part of the 1998/1999 outburst
of XTE J1550-564 (from \cite{homa2001}). Panel A and B are for different
gain epochs of the PCA detector. Open symbols, squares and triangles, 
mark the position of disk-blackbody and power law models respectively.
}\label{fig:homan3}
\end{figure*}

A complete spectral analysis of these data can be found in \cite{sobc2000}.
However, the second part of this outburst (after MJD 51139, 
see Fig. \ref{fig:sobczak6}) was
analyzed in terms of color-color diagrams (CCD) by \cite{homa2001},
showing in a model-independent way the spectral evolution of the system.
In Fig. \ref{fig:homan3} the CCD of 
XTE J1550-564 from \cite{homa2001} is shown.
The colors are defined in such a way that the diagram preserves linearity,
meaning that the sum of two models will result in a position lying between
the points characterizing those models. The position of disk-blackbody models
with different temperatures and power-law models with different photon
indices are shown. Notice that the disk-blackbody points lie on a roughly
vertical line, while the power-law line is much more inclined. This means that
changes in the parameters of the two models are almost decoupled in this
diagram.  In Fig. \ref{fig:homan3}, panel B refers to the
times after MJD 51259, when the PCA gain was changed. Looking at panel A,
we can follow the evolution the spectral distribution
and, assuming the spectrum can be approximated with the sum
of these models, we can have an idea of the state transitions.

At the beginning of the outburst the points are away from the DBB line,
indicating the presence of a substantial amount of power law. 
Being the total flux relatively low, we can tentatively call this
an IS. Then the source moves very close to the DBB line and slowly
moves up, increasing the temperature of the DBB component: a HS. What follows
is the flaring which can be seen in the HEXTE light curve in 
Fig. \ref{fig:sobczak6},
corresponding to large deviations from the DBB line: VHS excursions.
Overall, the evolution can be deconvolved in this way: the temperature of the
soft component first increases then decreases smoothly, moving the source 
firs up then down in the diagram. The power law component changes in 
intensity in a rather uncorrelated way, bringing first the source close
to the DBB line (small power-law contribution), then moving it away 
repeatedly during the flares.
Looking at the timing properties, we see a PDS
typical of a HS when the source is close to the DBB line, and extremely
complex power density spectra, dominated by 
multi-peaked QPOs, when the contribution
of the power law increases, strengthening the state identification.
At the very beginning and at the very end of the outburst, when the 
power-law component dominates the flux, the PDS is characteristic of the
LS \cite{homa2001,cui1999}.

\begin{figure}[t] 
\vspace{10pt}
\centerline{\psfig{file=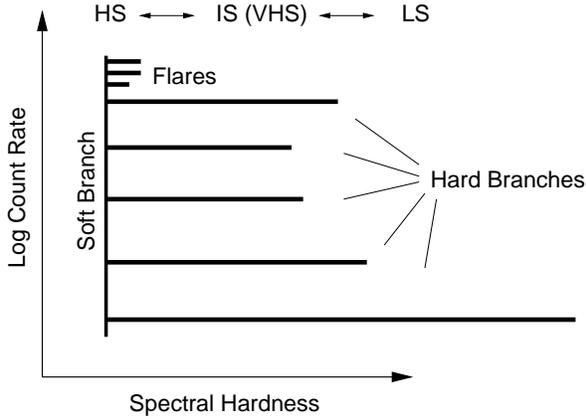,width=3.0in}}
\caption{Diagram of the state behavior of XTE J1550-564
from \cite{sobc2000}.
}\label{fig:homan26}
\end{figure}

What we have seen above is not what the simple picture of state transitions
driven by accretion rate would predict. There are state transitions, 
but they seem to be decoupled from accretion rate. 
Figure \ref{fig:homan26} shows a
schematic description of this (from \cite{homa2001}). While changes in the 
accretion rate are naturally responsible for variations in the temperature
of the soft component (corresponding to large changes in detected flux),
the power law component moves along different branches and transitions 
to the VHS can happen at different values of the accretion rate.
On the contrary, the LS appears only at the beginning and at the end
of the outburst, suggesting a connection to low accretion rates.

\section{Spectral/timing states II}

What shown above has important consequences for the picture of spectral/timing
states. The first, already mentioned, is that state transitions in BHC
are not driven {\it only} by mass accretion rate: a second independent 
parameter is needed. It is not in the scope of this paper to discuss physical
possibilities for what this second parameter is. In GS 1124-68, the
transitions were relatively simple and a sequence of states was identified
\cite{ebis1994}, while here things behaved in a more complex fashion.
Other sources, like GS 2023+338 \cite{tana1995}, 
remained in the LS despite large changes
in accretion rate, indicating that this second parameter did not vary. On
the contrary, in 
the 1996 transition of Cyg X-1, a relatively small change in accretion rate
was sufficient to trigger a state transition \cite{zhan1997}.
The second consequence is that the IS becomes indistinguishable from the VHS,
since its spectral/timing properties were identical to those of the VHS, the
difference being only in the range of accretion rate at which it appeared
\cite{bell1997,mend1997}.
If the VHS can be triggered at any accetion rate, no IS is needed.
This brings the number of states of black hole candidates back to {\bf three}.

\section{4U 1630-47}

\begin{figure}[t] 
\vspace{10pt}
\centerline{\psfig{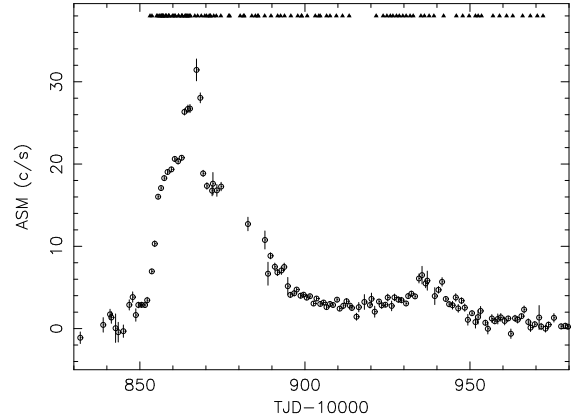}}
\caption{RXTE/ASM light curve of the 1998 outburst of 4U 1630-47. The
	triangles mark the dates of the PCA pointed observations. From 
	\cite{diet2000}
}\label{fig:dieters1a}
\end{figure}

\begin{figure}[!htb] 
\vspace{10pt}
\centerline{\psfig{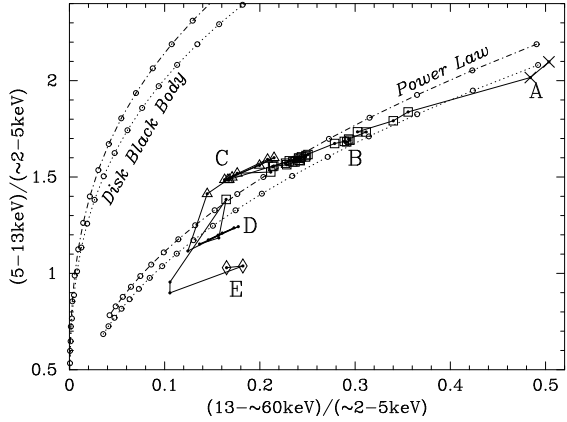}}
\caption{CCD of the 1998 outburst of 4U 1630-47. The outburst proceeds
	from A to D. From \cite{diet2000}.
}\label{fig:dieters1d}
\end{figure}

4U 1630-47 is a bright recurrent transient system that shows an outburst roughly
every 600 days (see \cite{kuul1997}). Since the launch of RXTE, it has showed
four outbursts; here I concentrate on the second one, during 1998.
The outburst light curve is shown in Fig. \ref{fig:dieters1a}, 
and the corresponding CCD
(defined in a similar way to that of XTE J1550-564 above) can be seen in 
Fig. \ref{fig:dieters1d} (both figures are from \cite{diet2000}).

In the case of 4U 1630-47, all points in the CCD lie close to the power-law
line. The fact that the source moves from below to above that line indicate
changes in the temperature of the soft component, but we never see a state
dominated by it like in the case of XTE J1550-564. This can also be seen
from the complete spectral analysis in \cite{trud2001}. The data from 
\cite{diet2000} indicate that the source was always in the VHS.
During the final stages of this outburst, the source entered the LS,
as expected (see \cite{toms2000}). A parallel timing analysis shows indeed
that the HS seems not to be entered at any time during this outburst. As in 
the case of XTE J1550-564, the VHS power spectra appear to be extremely
complex and difficult to interpret \cite{diet2000}.

\begin{figure}[!htb] 
\vspace{10pt}
\centerline{\psfig{file=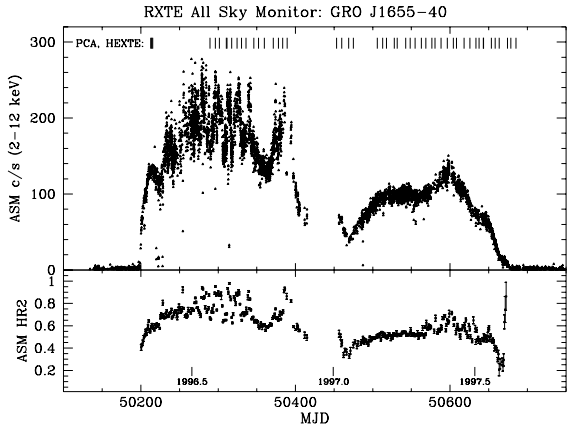,width=3.0in}}
\caption{RXTE/ASM light curve of the outburst of GRO J655-40 (from 
\cite{remi1999}). The bottom panel shows the corresponding ASM
	hardness ratio. The different states are indicated.
}\label{fig:remillard1}
\end{figure}

\begin{figure}[!htb] 
\vspace{10pt}
\centerline{\psfig{file=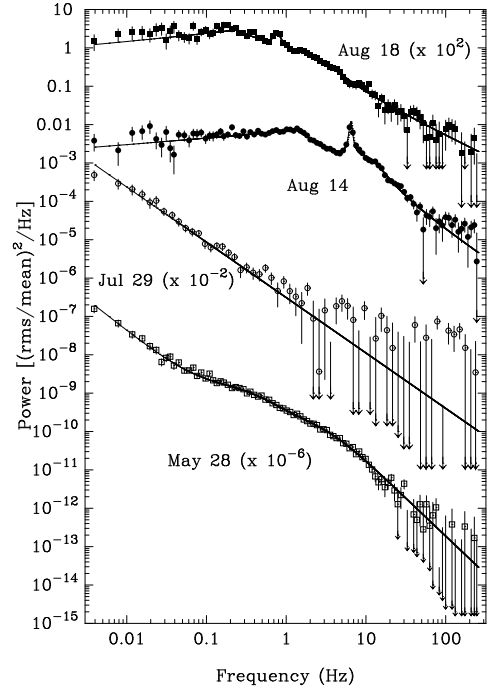,width=3.0in}}
\caption{Four Power spectra at the end of the outburst of GRO J1655-40.
	From \cite{mend1998}).
}\label{fig:mendez2}
\end{figure}

In the case of 4U 1630-47, it is difficult to extract from the CCD
information about the evolution of the mass accretion rate. However, 
five observations during this outburst were made with BeppoSAX during
the decay part of the outburst (between TJD 10864-10899): detailed 
spectral analysis with a DBB plus power law model indicate that the
data are consistent with no variation in the mass accretion rate 
\cite{oost1998}. Of course this result is strongly model-dependent, but
it suggests that rather large changes in the timing properties (see 
\cite{diet2000}) can appear without being driven by strong variations
in the mass accretion rate.

\section{GRO J1665-40}

GRO J1655-40 is one of the first two microquasars (see \cite{hjel1995}).
RXTE observed a major outburst of this source in 1996 
(see \cite{remi1999,sobc1999}). The RXTE/ASM light curve for this 
outburst is shown in Fig. \ref{fig:remillard1}. 
The outburst evolution looks similar to 
that of XTE J1550-564, with two separate parts, the second of which
looks smoother than the first. Spectrally, three states could be
identified: VHS in the first part of the outburst, HS in the second, 
with the exception of the last PCA observations, when the source switched
to the LS (\cite{sobc1999}). In the timing domain, things did not look
that simple. The only major difference between the first and the second
part of the outburst was the presence of QPO features in the former
\cite{remi1999}. You can see the overall shape of these PDS 
in Fig. \ref{fig:mendez2}, 
where four PDS at the end of the outburst are shown (from
\cite{mend1998}). The typical PDS for the source, QPOs excluded, is 
the bottom one, with a few \% fractional variability. In a month time
the source moved to a HS-like PDS and then to a typical PDS for a LS
(see Fig. \ref{fig:mendez2}).

From this, the question arises: how is the VHS defined in the
PDS domain? We have given a definition in Section 1, but after that we have
different and complex PDS which we classified as VHS.
This will be discussed later in this paper.

\begin{figure*}[!htb] 
\vspace{10pt}
{\psfig{file=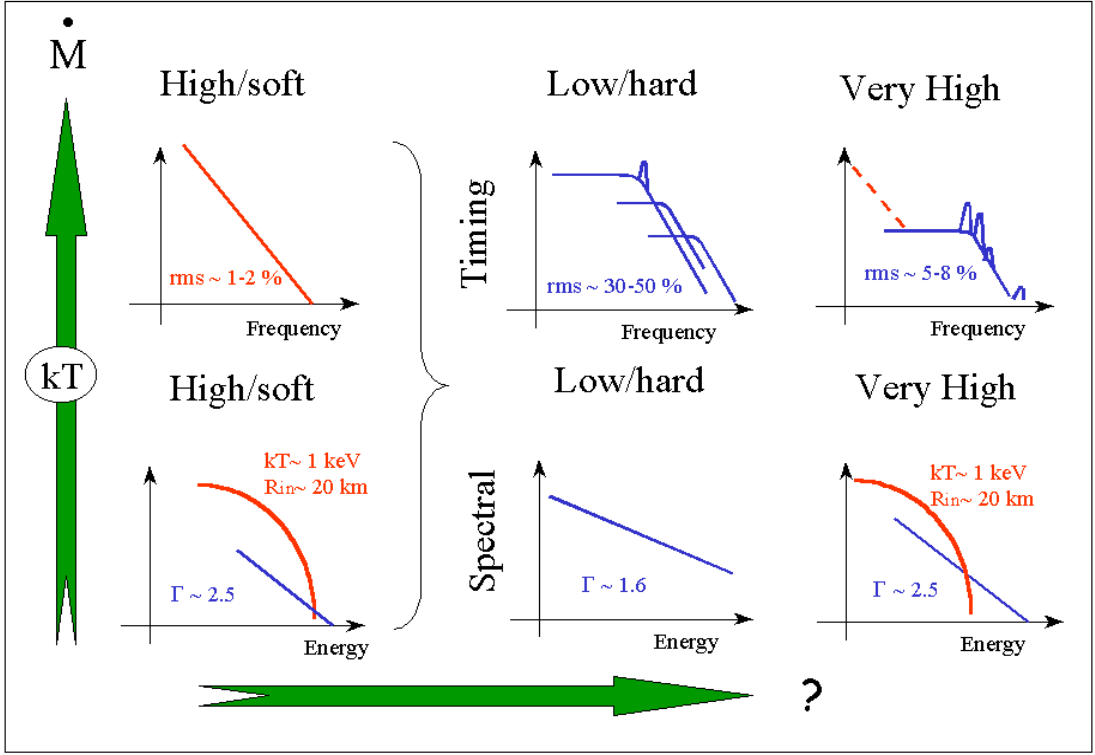,width=6.0in}}
\caption{Spectral/timing states and their dependence on mass accretion rate
	and the second (unknown) parameter.
}\label{fig:jhu}
\end{figure*}

\section{Spectral/timing states III}

From what we have seen, we can summarize a few points about states and
state-transitions:

\begin{figure*}[t] 
\vspace{10pt}
{\psfig{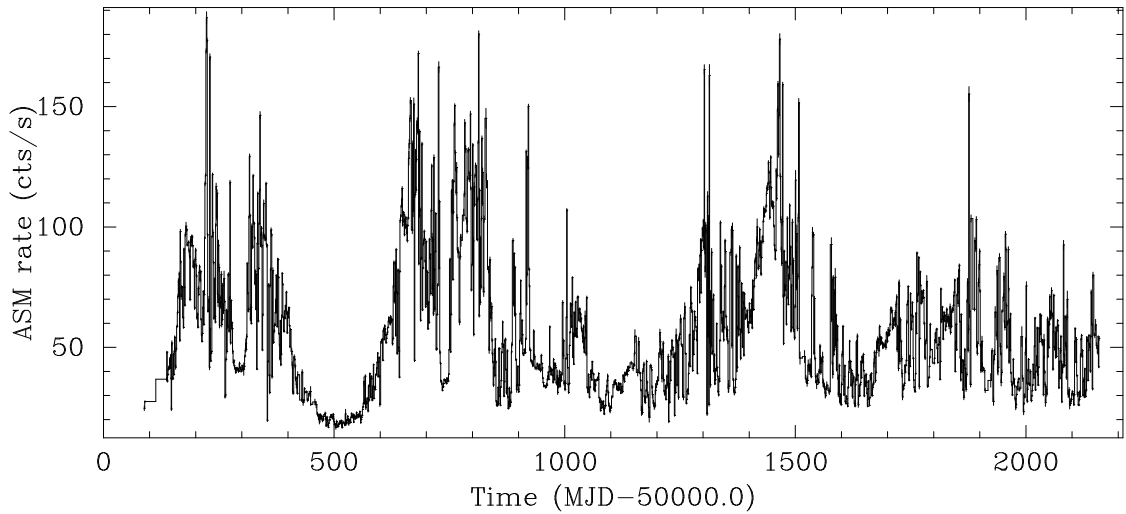}}
\caption{RXTE/ASM light curve of GRS 1915+105 (updated to 2001 September 6).
	Bin size is 1 day. 1 Crab $\sim$ 75 cts/s.
}\label{fig:asm1915}
\end{figure*}

\begin{itemize}

\item The soft disk component reacts to changes in the accretion rate, the
	hard component (a power law in the 2-20 keV band) reacts to something
	else (indicated as a question mark in Fig. \ref{fig:jhu}).

\item The LS is observed only at the beginning and at the end of the 
	outbursts, indicating its association with a low accretion rate

\item The main difference between VHS and HS is the presence of the hard
	component: if it is present, significant short-term variability
	is observed and the source is in the VHS. If it suppressed or
	absent, there is little variability and the source is in the HS.
	As indicated above, the appearance of the hard component is
	driven by a second parameter whose nature is not clear.

\item It must be noted that it is not at all clear that the power-law
	components observed in the VHS and in the HS are the same 
	component. In order to verify this, high-energy coverage is 
	necessary, to check for the presence/absence of a high-energy
	cutoff (see \cite{grov1998}).
	In the case of GRO J1655-40, orbit-related absorption dips
        were observed: during these dips, both the disk and the power-law
	component were heavily absorbed, and an additional weak steep
	power law was visible \cite{kuul1998}. It is possible that this
	component is the HS power law, which would only become detectable
	when the other much brighter hard component disappears.
\end{itemize}

\begin{figure*}[!htb] 
\vspace{10pt}
{\psfig{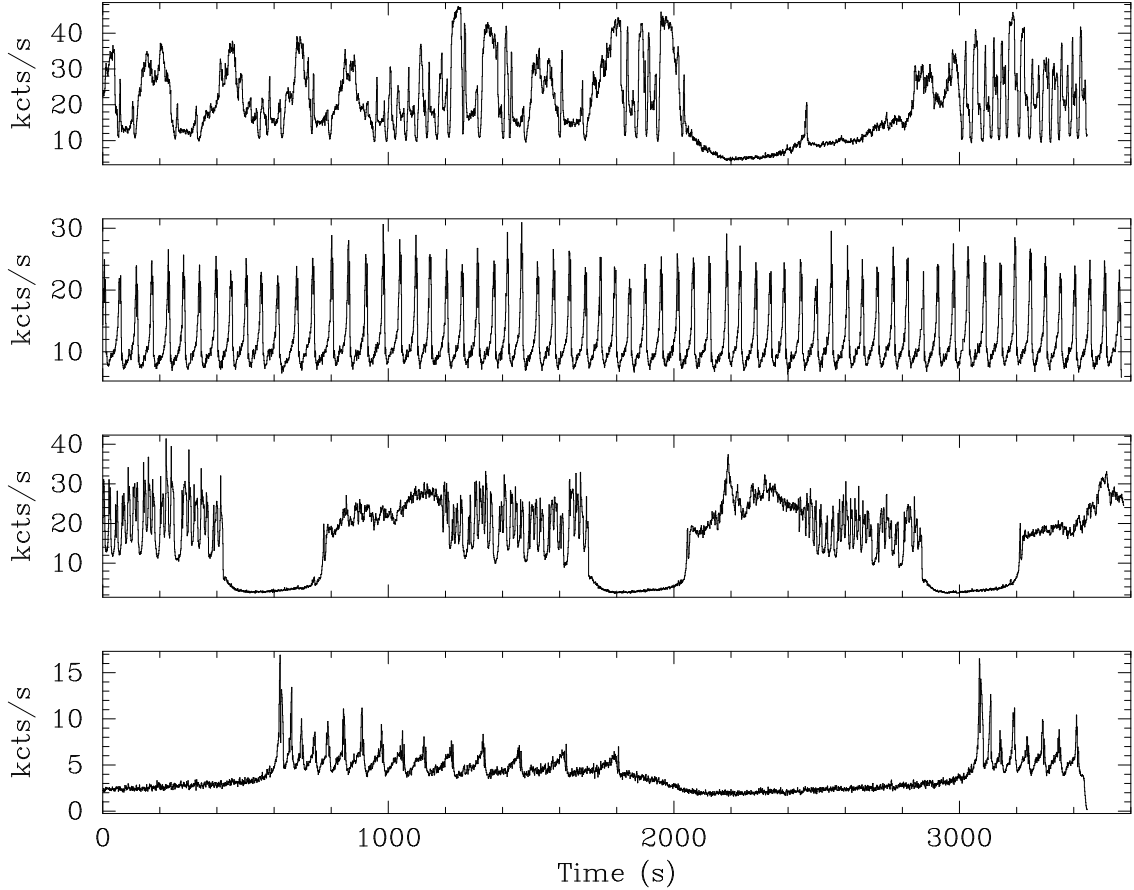}}
\caption{Four examples of one-hour RXTE/PCA light curves of GRS 1915+105.
	Bin size is 1 second.
}\label{fig:pca1915}
\end{figure*}

As mentioned above, the phenomenological description of the VHS needs to 
be discussed. We have seen very different PDS associated to this state,
which are much more complex than what shown in Fig. \ref{fig:jhu}. 
Indeed, it would
be hard to describe a typical PDS for the VHS. In the timing domain, the
difference with the HS is the presence of significant variability, whether
in the form of continuum components or QPOs, above the 1-2\% observed in 
the HS. This variability is associated to the hard component and not to 
the softer disk component, and indeed it is not observed when the hard
component is missing.

\section{GRS 1915+105}

The systems examined so far had properties that were not too different
from those mentioned in Sect. 1 for a well-behaved transient: their
outburst was a few months long, although more complex than a 
fast-rise/exponential decay, and they underwent state transitions
in the course of the outburst. The original microquasar, GRS 1915+105
did not behave equally well. As an X-ray source, it appeared in the 
sky in 1992 \cite{cast1992} and remained bright since then.

Figure \ref{fig:asm1915} shows the RXTE/ASM light curve of 
GRS 1915+105: it shows very 
marked variability, like the first part of the outburst of GRO J1655-40. 
However, unlike in the case of GRO J1655-40 (see \cite{remi1999}),
this variability extends to very low time scales 
\cite{grei1996,morg1997,bell2000}.  This can be seen in 
Fig. \ref{fig:pca1915}, where a few examples of 
RXTE/PCA light curves, binned at 1 second,
are shown.

\begin{figure}[!htb] 
\vspace{10pt}
\centerline{\psfig{file=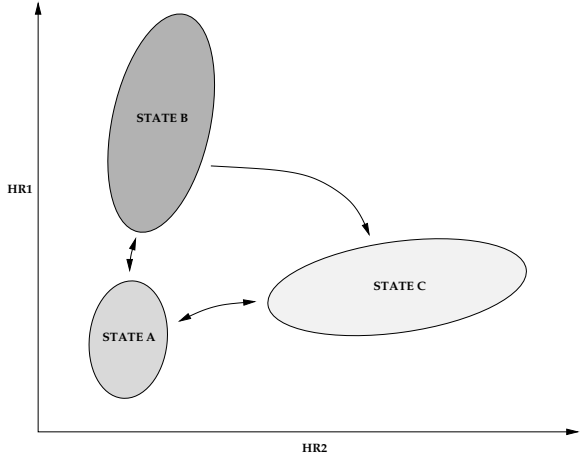,angle=-90,width=3.0in}}
\caption{Schematic color-color diagram showing the allowed transitions between
the three states of GRS 1915+105. From \cite{bell2000}.
}\label{fig:states}
\end{figure}

This extraordinary variability has been interpreted as due to the repeated
transitions between three basic states, called states
A, B and C \cite{bell2000}. A scheme of transitions between these states
is shown in Fig. \ref{fig:states} (from \cite{bell2000}). 
Is there a relation between these three states and the three canonical 
states described above? Of course, transitions as fast and repeated as those
shown in Fig. \ref{fig:pca1915} 
are not observed in any other source, but the similarities
might be stronger than the differences.
In order to answer, we need to examine the states
of GRS 1915+105 in more detail:

\begin{itemize}

\item {\it State B}: The energy spectrum consists of the superposition of 
	a strong soft thermal component (with kT$\sim$2 keV) and a
	steep hard component $\Gamma\geq$3. The PDS is steep, bumpy,
	with a total fractional variability in the 5-10\% range.

\item {\it State A}: The energy spectrum is similar to that of state B,
	but softer. This difference is attributed by \cite{bell2000} to
	a softer temperature of the thermal component. The PDS is
	a bumpy power law, with fractional rms lower than state B.

\item {\it State C}: The energy spectrum is dominated by the hard component,
	which is considerably flatter than in the other states, with
	$\Gamma$ that can be as low as 1.8. The PDS shows a band-limited
	noise component with a variable break frequency (between 0.07 Hz and
	a few Hz). A low-frequency QPO is always present, with a centroid
	frequency strongly correlated with count rate and spectral hardness
	\cite{reig2000}.

\end{itemize}

\begin{figure}[!htb] 
\vspace{10pt}
\centerline{\psfig{file=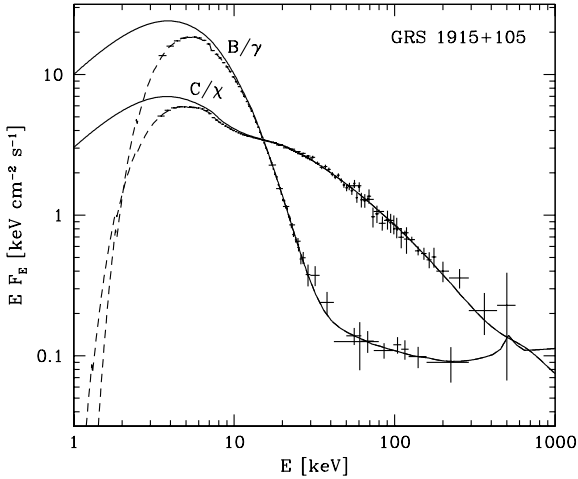,width=3.0in}}
\caption{Model spectra for the combined PCA/HEXTE/OSSE spectral fits of
	GRS 1915+105 in B and C state. No high-energy cutoff is
	needed for either spectra. From \cite{zdz2001}.
}\label{fig:zdz}
\end{figure}

As one can see, states A and B seem to have something in common with the
canonical VHS, while state C shows similarities with the LS, although 
the dependence of the timing parameters with rate and hardness 
is something that was seen before in the VHS of GS 1124-68 \cite{taki1997}.
In addition, by combining RXTE and CGRO/OSSE data, \cite{zdz2001} showed
that the broad-band spectrum of GRS 1915+105 in states C and B does
not show any evidence of a high-energy cutoff
(see Fig. \ref{fig:zdz}, unlike typical BHC in the
LS (see \cite{grov1998}).

At this stage, it is difficult (and dangerous) to claim any strong 
association  between the three states of GRS 1915+105 and the canonical
states of more conventional systems. The source is extremely bright and
possibly what we see is only intances of VHS at very high accretion rates.

\section{On the timing noise in the LS and in the VHS}

\begin{figure}[!htb] 
\vspace{10pt}
\centerline{\psfig{file=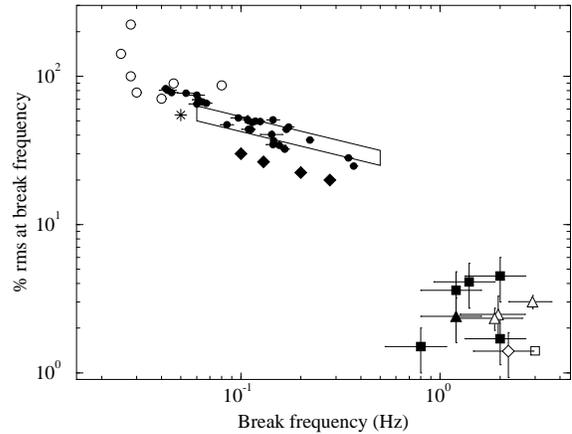,width=3.0in,bbllx=150pt,bblly=116pt,bburx=590pt,bbury=496pt}}
\caption{Flat-top level (in fractional rms) vs. break frequency for a
	sample of BHC. LS points are clustered in the upper left, VHS
	points in the lower right. From \cite{mend1997}.
}\label{fig:mendez339}
\end{figure}

As discussed above, there are indications that the hard components in the
LS and in the VHS are different, as a high-energy cutoff is detected in the
LS, but no high-energy cutoff is detected in the VHS \cite{grov1998}.
However, if one looks at the properties of the band-limited noise detected
in these two states, such as the correlation between flat-top and 
break frequency in the PDS, the VHS points lie on the high-frequency 
extrapolation of the LS ones (\cite{mend1997}, see Fig. \ref{fig:mendez339}).
This is difficult to understand if the two components have a different
physical origins: in this case both the characteristic noise frequency
and the total fractional rms (see \cite{bell2001}) must be associated 
either to geometrical parameters or to the thermal disk component.

\section{Spectral/timing states: conclusions}

Since the launch of RXTE, thanks to the presence of the ASM and the
operational flexibility of the mission, the available X-ray data on 
black hole candidates, especially transient systems, as increased
substantially. The picture that emerges from the analysis of this
large database is at the same time simpler and more complex than 
the paradigm that existed before. It is more complex because the
phenomenology became quite complicated. The variety of QPOs observed
in XTE J1550-564 and GRS 1915+105, and the extreme structured variability
of the latter are perhaps the best examples. However, it also
became somewhat simpler, as the number of basic spectral/timing states
is now reduced to three, despite the tremendous diversity of some
timing features, and the presence of a second parameter governing state
transitions (although its physical nature needs to be addressed) might
yield a direct measurement of a fundamental parameter of these systems.
Among all sources, GRS 1915+105 shows more variability and state-transitions
than all other sources together. it might be the way to solving basic
problems of accretion, but it might also turn out to be an endless 
complication that leads away from the solutions.
However, it is important that theoretical models address the general
picture described above in addition to trying to describe in extreme
detail the spectral distribution of a particular state in a particular
source, especially since any spectral fit to low-resolution data 
involves by definition a strong a priori bias. 


\normalsize

\section*{ACKNOWLEDGEMENTS}
I thank the Cariplo Foundation for financial support.
 
\end{document}